\documentclass[american,9pt,conference,a4paper]{IEEEtran}
\usepackage[latin9]{inputenc}
\usepackage{float}
\usepackage{booktabs}
\usepackage{mathtools}
\usepackage{amsmath}
\usepackage{amssymb}
\usepackage{graphicx}
\usepackage{esint}

\makeatletter

\providecommand{\tabularnewline}{\\}

\usepackage{fancyhdr}

\usepackage{pstricks, pst-node, pst-plot, pst-circ}
\usepackage{graphicx}

\usepackage{psfrag}

\usepackage{xcolor}
\usepackage{xfrac}

\usepackage{url}

\usepackage{amsmath}
\usepackage{amssymb}
\usepackage{array}

\usepackage{booktabs}

\usepackage{multirow}

\usepackage{times,epsfig}
\usepackage{epstopdf}
\usepackage{rotating}

\usepackage{pgfplots}
\pgfplotsset{compat=newest}
\pgfplotsset{plot coordinates/math parser=false}

\newcommand{\fse}{FSE}
\newcommand{\fsecentroid}{CA-FSE}
\newcommand{\fsecoordy}{m}
\newcommand{\fsecoordx}{n}
\newcommand{\fsew}{w\left[\fsecoordy,\fsecoordx\right]}
\newcommand{\fsemodel}{g}
\newcommand{\fsecoeff}{\hat{c}}
\newcommand{\fsebasefkt}{\varphi}
\newcommand{\fserho}{\rho}
\newcommand{\fseodc}{\gamma}
\newcommand{\fsedelta}{\delta}
\newcommand{\fsesizey}{M}
\newcommand{\fsesizex}{N}
\newcommand{\fsecoordycenter}{\bar{\fsecoordy}}
\newcommand{\fsecoordxcenter}{\bar{\fsecoordx}}
\newcommand{\fsecoordycentroid}{\fsecoordy_c}
\newcommand{\fsecoordxcentroid}{\fsecoordx_c}
\newcommand{\fseareaproc}{\mathcal{L}}
\newcommand{\fsearearec}{\mathfrak{b}}

\newcommand{\fsesupborder}{d}
\newcommand{\fsepixelorig}{\mathcal{A}}
\newcommand{\fsepixellost}{\mathcal{B}}
\newcommand{\fsepixelrec}{\mathcal{R}}

\newcommand{\blockbased}{block-based}

\newcommand{\spaceBeforeEqhalf}{}
\newcommand{\spaceBelowTab}{}

\newcommand{\spaceBelowFig}{}
\newcommand{\spaceBeforeLabel}{}
\newcommand{\spaceBeforeLabelhalf}{}

\newcommand{\fig}{Fig.}
\newcommand{\tab}{Tab.}

\newcommand{\Fig}[1]{\fig{}~#1}
\newcommand{\Tab}[1]{\tab{}~#1}

\newcommand{\CSALSA}{\mbox{C\hspace{-0.15mm}S\hspace{-0.15mm}A\hspace{-0.15mm}L\hspace{-0.15mm}S\hspace{-0.15mm}A}}

  \newlength\figureheight
    \newlength\figurewidth
\newcommand{\fakebuchstabe}{}

\author{
\vspace{2mm}\\
{Wolfgang Schnurrer, Markus Jonscher, Jürgen Seiler, Thomas Richter, Michel Bätz, and André Kaup}
\vspace{2mm}\\ 
\fontsize{10}{10}\selectfont\itshape 
Multimedia Communications and Signal Processing\\
Friedrich-Alexander Universität Erlangen-Nürnberg (FAU)\\
Cauerstr. 7, 91058 Erlangen, Germany
\fontsize{9}{9}\selectfont\ttfamily\upshape

\vspace{1.5mm}\\ 
\fontsize{10}{10}\selectfont\rmfamily\itshape 

\fontsize{9}{9}\selectfont\ttfamily\upshape 
Email: \{ schnurrer, jonscher, seiler, richter, baetz, kaup \} @lnt.de\\
\vspace{2mm}
}

\makeatother

\usepackage{babel}
\begin{document}

\title{Centroid Adapted Frequency Selective Extrapolation \\
for Reconstruction of Lost Image Areas}

\maketitle
\begin{figure}[b] 
\parbox{\hsize}{\em 
}\end{figure}
\begin{abstract}
Lost image areas with different size and arbitrary shape can occur
in many scenarios such as error-prone communication, depth-based image
rendering or motion compensated wavelet lifting. The goal of image
reconstruction is to restore these lost image areas as close to the
original as possible. Frequency selective extrapolation is a block-based
method for efficiently reconstructing lost areas in images. So far,
the actual shape of the lost area is not considered directly. We propose
a centroid adaption to enhance the existing frequency selective extrapolation
algorithm that takes the shape of lost areas into account. To enlarge
the test set for evaluation we further propose a method to generate
arbitrarily shaped lost areas. On our large test set, we obtain an
average reconstruction gain of 1.29~dB. 
\\[1\baselineskip]
\end{abstract}


\section{Introduction}

Image reconstruction aims at restoring lost areas in images as close
as possible to the original. There are several applications where
lost areas of arbitrary shape can occur and need to be reconstructed,
e.g., when distortions like scratches or dust are to be removed from
scanned images. In multiview scenarios, lost areas can occur especially
at object boundaries, when an intermediate view is computed by depth-image
based rendering \cite{fehn2004dibr}. In motion compensated frame
rate up conversion \cite{sunwoo2014}, lost areas can occur in predicted
intermediate frames. A very similar pattern of lost areas occurs when
the block-based motion compensation is inverted, e.g., in the update
step of compensated wavelet lifting, unconnected pixels can occur
\cite{bozinovic2005}. In \cite{schnurrer2013}, the reconstruction
of these unconnected pixels was shown to be advantageous. In all
of these applications, the different size and the arbitrary shape
of the occurring lost areas is challenging.

Several methods exist for reconstructing lost image areas. In~\cite{kernelregression2007},
classical kernel regression (CKR) is extended by nonlinear kernel
adaption to obtain steering kernel regression (SKR). \cite{totalvariation2010}
proposes a framework based on total variation (TV) that can be used
for reconstructing larger lost areas in images, e.g., to remove text
overlays. In~\cite{salsa2011}, the constraint split augmented Lagrangian
shrinkage algorithm (\CSALSA{}) is proposed and used for reconstructing
images with lost pixels. The frequency selective extrapolation (\fse{})
generates a model in the frequency domain based on the available pixels
\cite{seiler2010}. 

In natural images, pixels closer to each other have a higher correlation.
The correlation reduces with increasing distance of the pixels. When
some pixels are lost, support pixels closer to these lost pixels should
have a higher influence on the reconstruction. \fse{} is a \blockbased{}
method which uses an isotropic weighting function to control the influence
of the support pixels. So far, mostly block losses were considered
where the currently considered block was completely lost, i.e., the
weighting function is centered w.r.t. the currently considered block.

\begin{figure}
 \setlength\tabcolsep{2pt}

\begin{center}

\begin{tabular}{ccccc}
\includegraphics[height=3.2cm]{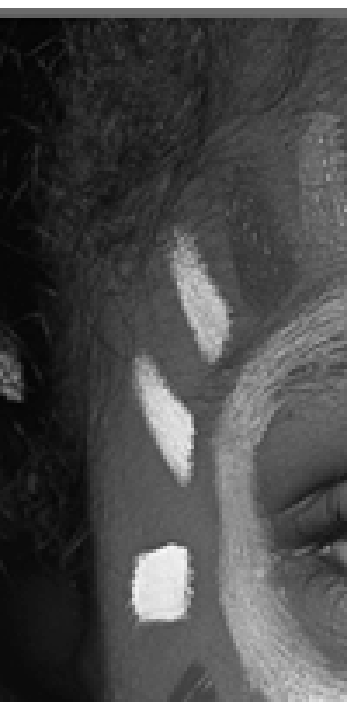} & ~\includegraphics[height=3.2cm]{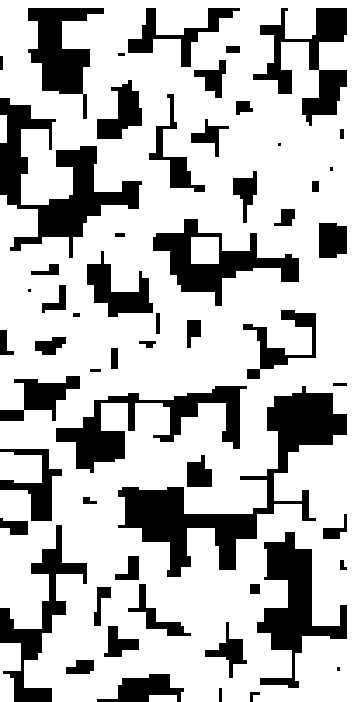} & \includegraphics[height=3.2cm]{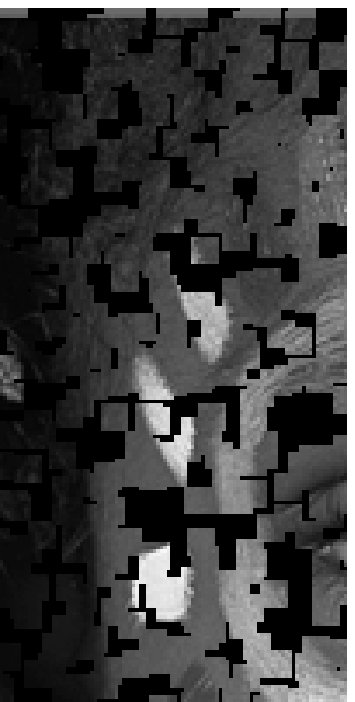} & ~\includegraphics[height=3.2cm]{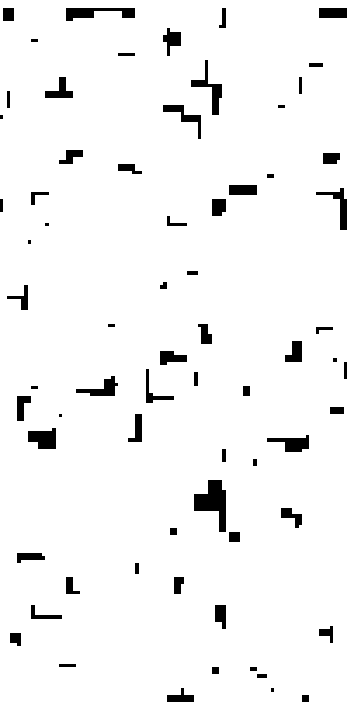} & \includegraphics[height=3.2cm]{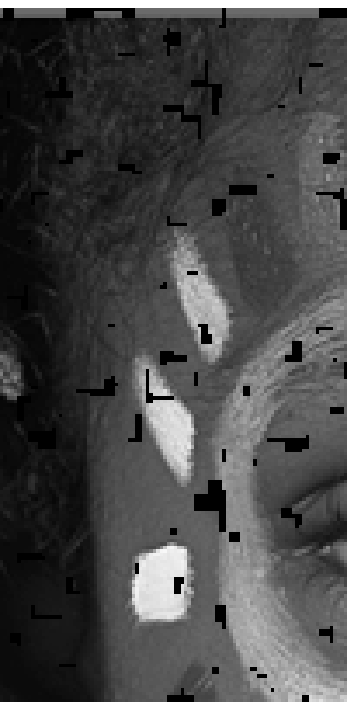}{\footnotesize{}}\tabularnewline
{\footnotesize{}Original} & \multicolumn{2}{c}{{\footnotesize{}Dense loss pattern}} & \multicolumn{2}{c}{{\footnotesize{}Sparse loss pattern}}\tabularnewline
\end{tabular}

\end{center}

\spaceBeforeLabelhalf{\footnotesize{}\vspace{-2mm}}

\protect\caption{\label{fig:Eyecatcher}Detail of an image, arbitrarily shaped loss
patterns (black is lost), and images to be reconstructed}

\spaceBelowFig\spaceBelowFig\spaceBelowFig\spaceBelowFig\spaceBelowFig
\end{figure}

In this paper, we focus on the reconstruction of images with arbitrarily
shaped lost areas. We propose a centroid adaption of \fse{} (\fsecentroid{})
to address the arbitrary shape of lost areas.  To evaluate the performance,
we propose a method to simulate the arbitrarily shaped lost areas
occurring in the above mentioned scenarios. \Fig{\ref{fig:Eyecatcher}}
shows an example image with dense and sparse loss pattern. By applying
different loss patterns to different images, the performance of the
reconstruction methods is evaluated and compared on a very large test
set.

In Section~\ref{sec:FSE}, we briefly review the \fse{} algorithm
and introduce our proposed centroid adaption in Section~\ref{sec:Centroid-Adaption}.
Simulation results and discussion follow in Section~\ref{sec:Simulation-Results}.

\section{Frequency Selective Extrapolation}

\label{sec:FSE}Frequency selective extrapolation (\fse{}) \cite{seiler2010}
is a block-based iterative method for reconstructing lost areas in
images. With the optimized processing order \cite{seiler2011},
the size of lost areas is taken into account. The more available support
pixels a block has, the earlier it is processed and bigger lost areas
are processed from the border to the center. However, in contrast
to the size, the shape of the lost areas is not considered so far.

For every block,  \fse{} generates a model\spaceBeforeEqhalf
\[
\fsemodel\left[\fsecoordy,\fsecoordx\right]=\sum_{k\in\mathcal{K}}\fsecoeff_{k}\fsebasefkt_{k}\left[\fsecoordy,\fsecoordx\right]
\]
for the unknown pixels based on the available pixels in the support
area. A weighted superposition of 2-D Fourier basis functions $\fsebasefkt_{k}$
is generated where the set $\mathcal{K}$ contains the indexes of
all basis functions. In every iteration, the influence $\fsecoeff_{k}$
of the basis function $\fsebasefkt_{k}$ is increased that reduces
the approximation error the most.

\begin{figure}
\psfragscanon

\psfrag{tag0}{Pixel categories of $\fseareaproc$:}

\psfrag{tag1}{originally known $\fsepixelorig$ }

\psfrag{tag2}{already reconstructed $\fsepixelrec$}

\psfrag{tag3}{$\fsepixellost_{o}$: lost, outside of $\fsearearec$}

\psfrag{tag4}{$\fsepixellost_{i}$: lost, inside of $\fsearearec$}

\psfrag{tag5}{$\fseareaproc=\fsepixelorig\cup\fsepixelrec\cup\fsepixellost_{i}\cup\fsepixellost_{o}$}

\psfrag{m}{$\fsecoordy$}

\psfrag{n}{$\fsecoordx$}

\psfrag{b}{$\fsearearec$}

\psfrag{d}{$\fsesupborder$}

\includegraphics[width=0.48\columnwidth]{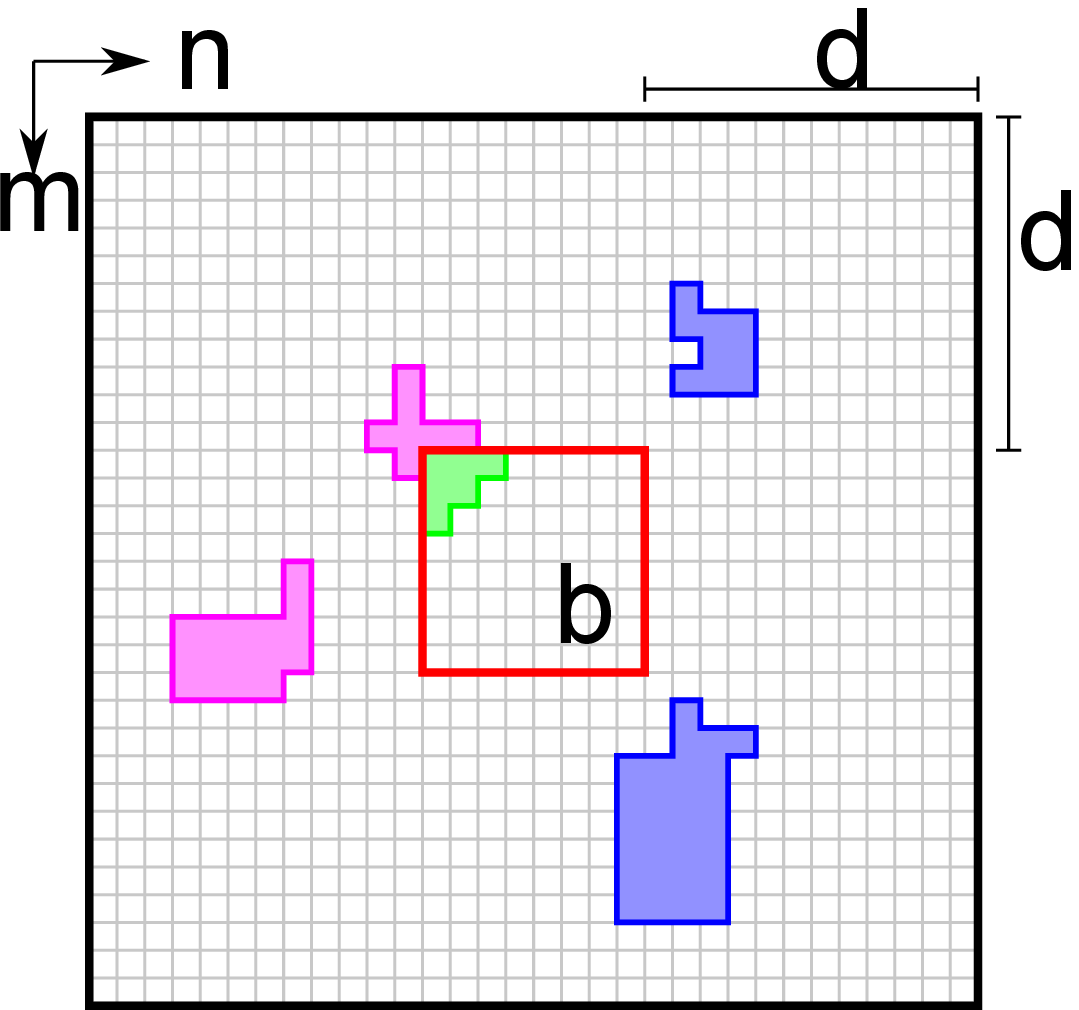}\includegraphics[width=0.48\columnwidth]{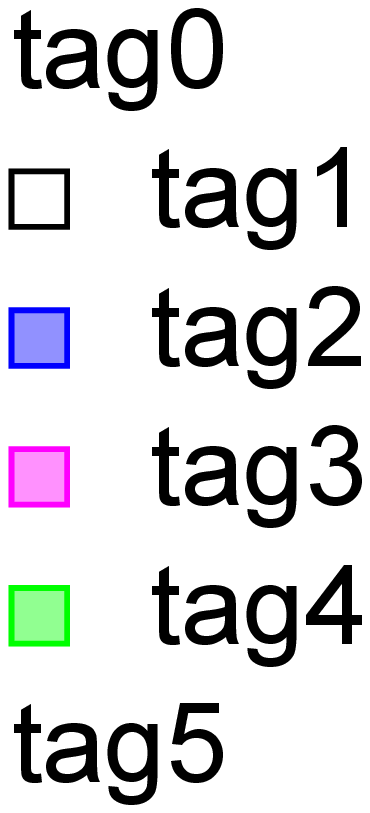}

\psfragscanoff

\spaceBeforeLabelhalf

\protect\caption{\label{fig:Composition-of-extrapolation}Composition of extrapolation
area $\fseareaproc$. The corresponding weighting functions are shown
in \Fig{\ref{fig:example-w}}.}

\spaceBelowFig\spaceBelowFig\spaceBelowFig\spaceBelowFig\spaceBelowFig
\end{figure}

\Fig{\ref{fig:Composition-of-extrapolation}} shows the composition
of the extrapolation area $\fseareaproc$. The area $\fseareaproc$
has a size of $\fsesizey\times\fsesizex$ pixels and consists of the
currently considered block~$\fsearearec$ in the center, surrounded
by a support area of width $\fsesupborder$. The pixels within $\fseareaproc$
come from three categories, namely originally known pixels $\fsepixelorig$,
already reconstructed pixels $\fsepixelrec$  and lost pixels $\fsepixellost_{i}\cup\fsepixellost_{o}$
that have not been reconstructed so far. Thereby, $\fsepixellost_{i}$
contains lost pixels within the currently considered block~$\fsearearec$
and $\fsepixellost_{o}$ contains lost pixels outside of $\fsearearec$,
respectively. Lost pixels within the currently considered block~$\fsearearec$
are reconstructed by an inverse transform of the model $\fsemodel$.
For a more detailed description of \fse{} together with pseudo code,
please refer to \cite{seiler2010,seiler2011}.

\section{Proposed Centroid Adaption}

To control the influence of the support pixels during the model generation,
a weighting function $\fsew$ is used. The influence decreases with
increasing distance to the center of $\fsearearec$. The weighting
function is computed to

\begin{figure}
\includegraphics[width=0.48\columnwidth]{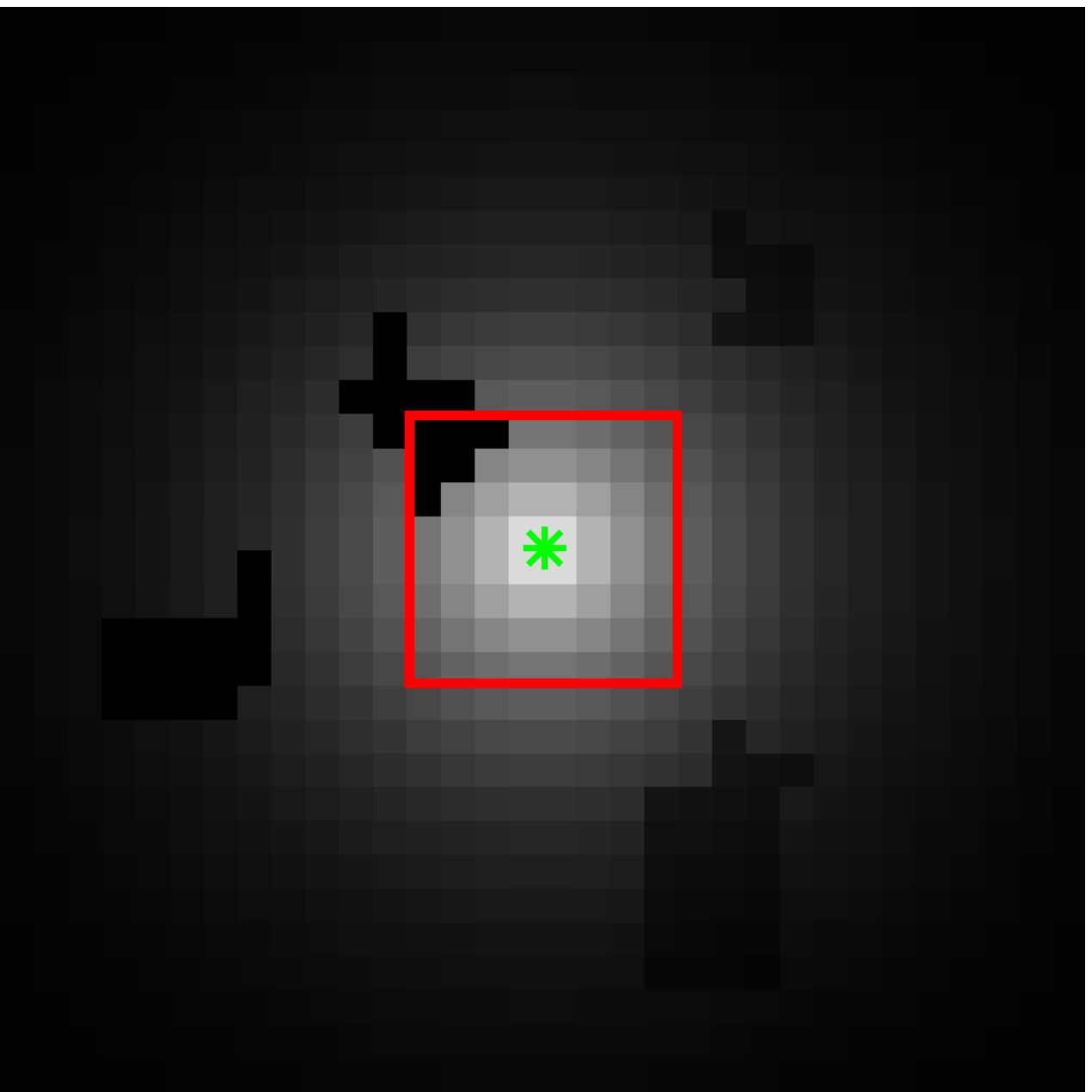}\hfill{}\includegraphics[width=0.48\columnwidth]{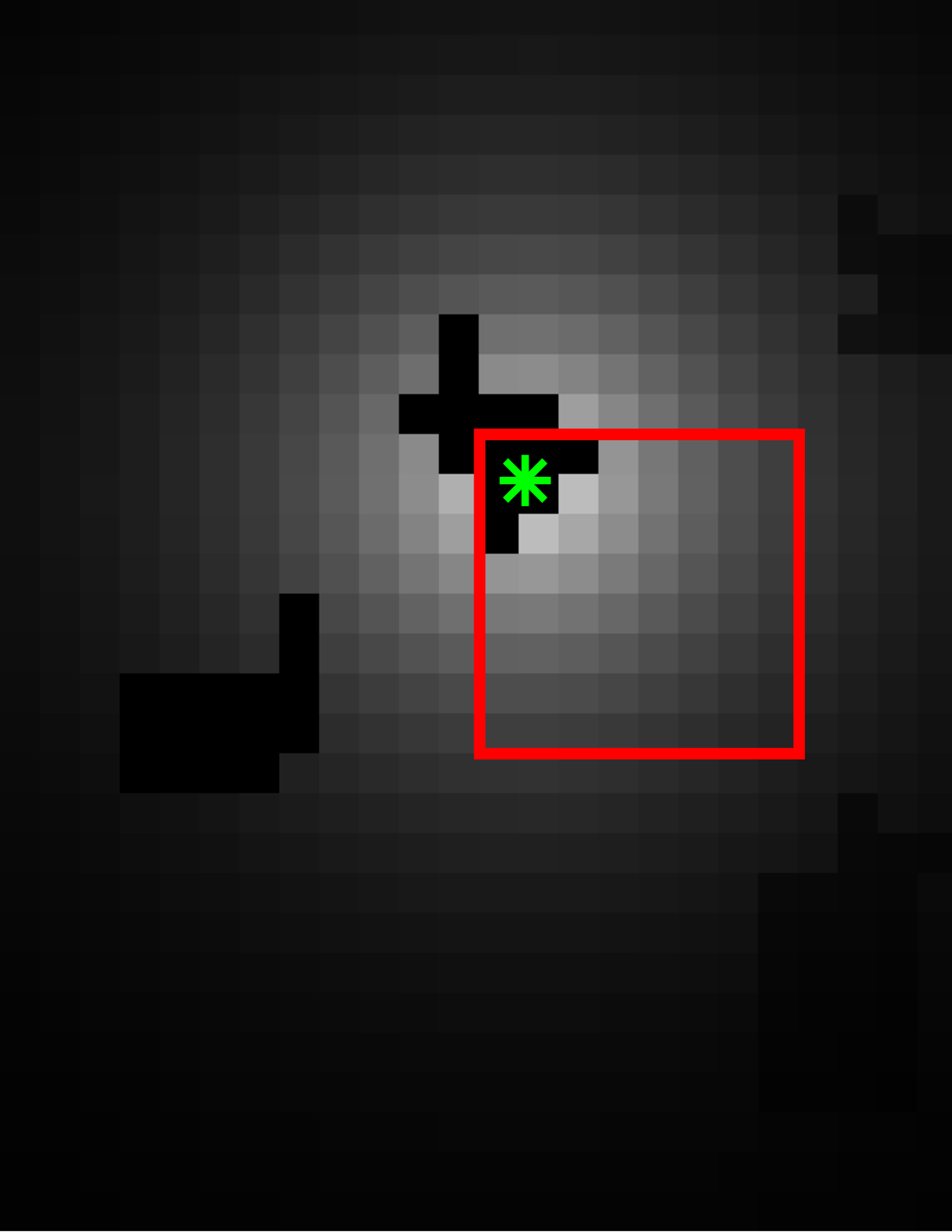}

\spaceBeforeLabelhalf

\protect\caption{\label{fig:example-w}Example for moving the center (marked by a green
star) of the weighting function $\fsew$ to the centroid of the lost
pixels (black) within the currently considered block $\fsearearec$
(red box). The different kinds of pixels are labeled in \Fig{\ref{fig:Composition-of-extrapolation}}.
Left: $\fsew$ centered in the currently considered block $\fsearearec$,
Right: $\fsew$ centered on the centroid of the lost pixels inside
the currently considered $\fsearearec$}

\spaceBelowFig\spaceBelowFig\spaceBelowFig\spaceBelowFig\spaceBelowFig
\end{figure}

\spaceBeforeEqhalf{}
\begin{equation}
\fsew\hspace{-1mm}=\hspace{-1mm}\begin{cases}
\fserho^{\sqrt{\left(\fsecoordy-\fsecoordycenter\right)^{2}+\left(\fsecoordx-\fsecoordxcenter\right)^{2}}} & \hspace{-1mm}\hspace{-1mm}\text{for }\hspace{-1mm}\left[\fsecoordy,\fsecoordx\right]\hspace{-0.5mm}\in\hspace{-0.5mm}\fsepixelorig\\
\fsedelta\fserho^{\sqrt{\left(\fsecoordy-\fsecoordycenter\right)^{2}+\left(\fsecoordx-\fsecoordxcenter\right)^{2}}} & \hspace{-1mm}\hspace{-1mm}\text{for }\hspace{-1mm}\left[\fsecoordy,\fsecoordx\right]\hspace{-0.5mm}\in\hspace{-0.5mm}\fsepixelrec\\
0 & \hspace{-1mm}\hspace{-1mm}\text{for }\hspace{-1mm}\left[\fsecoordy,\fsecoordx\right]\hspace{-0.5mm}\in\hspace{-0.5mm}\fsepixellost_{i}\hspace{-0.5mm}\cup\hspace{-0.5mm}\fsepixellost_{o}
\end{cases}\label{eq:fseweight}
\end{equation}
where $\left[\fsecoordycenter,\fsecoordxcenter\right]$ is the center
of the block~$\fsearearec$. On the left of \Fig{\ref{fig:example-w}},
an example is shown with $\fsew$ centered in $\fsearearec$ corresponding
to the areas named in \Fig{\ref{fig:Composition-of-extrapolation}}.
The decay is controlled by $\fserho$ and the weight of already reconstructed
pixels $\fsepixelrec$ is attenuated by $\fsedelta$.

So far, mostly block losses were considered where $\fsearearec$ was
completely lost. Especially for arbitrarily shaped lost areas, the
case occurs quite often where only a part of the reconstruction area
is actually lost.

\label{sec:Centroid-Adaption}The weighting function $\fsew$ controls
the influence of the support pixels on the model generation. Support
pixels closer to the lost area shall have more influence than pixels
farther away. So far, the weighting function does not consider the
case that only a part of the pixels within $\fsearearec$ can be lost.
To adapt the model generation to the arbitrary shape of the lost area,
we propose to center the weighting function on the centroid of the
lost pixels $\fsepixellost_{i}$, i.e., within $\fsearearec$. The
proposed weighting function is computed by

\spaceBeforeEqhalf{}
\begin{equation}
\fsew\hspace{-1mm}=\hspace{-1mm}\begin{cases}
\fserho^{\sqrt{\left(\fsecoordy-\fsecoordycentroid\right)^{2}+\left(\fsecoordx-\fsecoordxcentroid\right)^{2}}} & \hspace{-1mm}\hspace{-1mm}\text{for }\hspace{-1mm}\left[\fsecoordy,\fsecoordx\right]\hspace{-0.5mm}\in\hspace{-0.5mm}\fsepixelorig\\
\fsedelta\fserho^{\sqrt{\left(\fsecoordy-\fsecoordycentroid\right)^{2}+\left(\fsecoordx-\fsecoordxcentroid\right)^{2}}} & \hspace{-1mm}\hspace{-1mm}\text{for }\hspace{-1mm}\left[\fsecoordy,\fsecoordx\right]\hspace{-0.5mm}\in\hspace{-0.5mm}\fsepixelrec\\
0 & \hspace{-1mm}\hspace{-1mm}\text{for \hspace{-1mm}}\left[\fsecoordy,\fsecoordx\right]\hspace{-0.5mm}\in\hspace{-0.5mm}\fsepixellost_{i}\hspace{-0.5mm}\cup\hspace{-0.5mm}\fsepixellost_{o}
\end{cases}\label{eq:fseweight-centroid}
\end{equation}
where the centroid $\left[\fsecoordycentroid,\fsecoordxcentroid\right]$
of the lost pixels $\fsepixellost_{i}$ within the block~$\fsearearec$
is computed by

\spaceBeforeEqhalf{}
\begin{eqnarray*}
\fsecoordycentroid=\frac{1}{\left|\fsepixellost_{i}\right|}\sum_{\forall\fsecoordy\in\fsepixellost_{i}}\fsecoordy &  & \fsecoordxcentroid=\frac{1}{\left|\fsepixellost_{i}\right|}\sum_{\forall\fsecoordx\in\fsepixellost_{i}}\fsecoordx
\end{eqnarray*}

On the right of \Fig{\ref{fig:example-w}}, an example for centering
$\fsew$ on the centroid of the lost pixels within $\fsearearec$
is shown. This leads to the intended weighting of the pixels with
respect to the distance of the actual lost area. In the following,
the centroid adapted \fse{} is called \fsecentroid{}.

\section{Simulation Results and Analysis}

\label{sec:Simulation-Results}For evaluation, we used the images
from the two image databases \textit{Tecnick} \cite{tecnick}, consisting
of 100 images of size $1200\times1200$, and \textit{Kodak} \cite{kodak},
consisting of 24 images of size $512\times768$. To extend our test
set, we took the first image of each sequence of the \textit{ARRI}
database \cite{arri2013}. We further took views 0 from Illumination
1 and Exposure 1 of the \textit{Middleburry} multiview databases from
2005, 2006, and 2014 \cite{middleburry2007a,middleburry2007b,middleburry2014}.
The luminance of each image was used.

We used three typical parameter sets for \fse{} and \fsecentroid{},
called `bs 4', `bs 8', and `bs 16' listed in \Tab{\ref{tab:fse-param}}.

\begin{table}[H]
\spaceBeforeLabelhalf

\begin{center}

\protect\caption{\label{tab:fse-param}FSE parameter sets.}

\spaceBeforeLabelhalf

 \renewcommand\arraystretch{0.5} %
\begin{tabular}{cccc}
\toprule 
FSE parameter set denoted as & bs 4 &  bs 8 & bs 16\tabularnewline
\midrule
size of block $\fsearearec$ & $4\times4$ & $8\times8$ & $16\times16$\tabularnewline
border width $\fsesupborder$ around $\fsearearec$ & 14 & 12 & 16\tabularnewline
FFT size & 32 & 32 & 64\tabularnewline
decay parameter for $\fsew$ & \multicolumn{3}{c}{$\fserho=0.7$}\tabularnewline
attenuated weight for  $\fsepixelrec$ & \multicolumn{3}{c}{$\fsedelta=0.5$}\tabularnewline
orthogonality correction $\fseodc$ & \multicolumn{3}{c}{$\fseodc=0.5$}\tabularnewline
\bottomrule
\end{tabular}\end{center}

\spaceBelowTab\spaceBelowTab\spaceBelowTab
\end{table}

\begin{figure*}
%
%
\definecolor{mycolor1}{rgb}{0.07638,0.49828,0.84134}%
\definecolor{mycolor2}{rgb}{0.98736,0.73842,0.25149}%
\definecolor{mycolor3}{rgb}{0.38188,0.74505,0.53073}%
\begin{tikzpicture}

\begin{axis}[%
width=0.95092\figurewidth,
height=\figureheight,
at={(0\figurewidth,0\figureheight)},
scale only axis,
xmin=0,
xmax=25,
xlabel={\textit{Kodak} image number},
ymin=0,
ymax=3.5,
ylabel={$\text{PSNR}_{\text{\fsecentroid}} - \text{PSNR}_{\text{\fse}}$ in dB},
axis x line*=bottom,
axis y line*=left,
legend style={legend cell align=left,align=left,draw=white!15!black}
]
\addplot [color=mycolor1,solid,line width=1.5pt,mark=x,mark options={solid}]
  table[row sep=crcr]{%
1	0.586982944680937\\
2	0.619253985543754\\
3	1.46218865778276\\
4	1.15253090516348\\
5	0.544184646250681\\
6	1.27901399086836\\
7	0.745656157665692\\
8	0.716235778666881\\
9	0.893138732674942\\
10	1.07758533546804\\
11	0.521659050248264\\
12	2.76907696648258\\
13	0.440094082069606\\
14	0.7256571152204\\
15	3.1490895781137\\
16	0.848177917014304\\
17	1.40198335125882\\
18	0.432770110437989\\
19	0.506633245437655\\
20	1.19766544774958\\
21	0.463064634175552\\
22	0.379211873958461\\
23	0.563110429400325\\
24	0.580641212579078\\
};
\addlegendentry{bs 16};

\addplot [color=mycolor2,solid,line width=1.5pt,mark=triangle,mark options={solid}]
  table[row sep=crcr]{%
1	0.228268813049159\\
2	0.295830144466347\\
3	0.803931182623046\\
4	0.580949342004232\\
5	0.208942100311006\\
6	0.557235258946726\\
7	0.33375113838007\\
8	0.333155556268874\\
9	0.373581060605815\\
10	0.53334335821005\\
11	0.219735906038444\\
12	1.29466915994998\\
13	0.224551551372748\\
14	0.351303821273447\\
15	1.53341298624339\\
16	0.429141766979473\\
17	0.647849830058767\\
18	0.16532822702186\\
19	0.225755666831621\\
20	0.728144483965419\\
21	0.163090182560722\\
22	0.162173721931399\\
23	0.174028601041925\\
24	0.212621620163947\\
};
\addlegendentry{bs 8};

\addplot [color=mycolor3,solid,line width=1.5pt,mark=o,mark options={solid}]
  table[row sep=crcr]{%
1	0.153000701253184\\
2	0.216049845088644\\
3	0.72349370974975\\
4	0.503252201097251\\
5	0.114919227003931\\
6	0.42742885740839\\
7	0.207714773955711\\
8	0.243573770897459\\
9	0.244701276791254\\
10	0.493060039575596\\
11	0.117238475605202\\
12	1.14274053575611\\
13	0.162481636809236\\
14	0.245957610819286\\
15	1.26290185078147\\
16	0.3300301318761\\
17	0.496613522111257\\
18	0.10418165887172\\
19	0.146849880846123\\
20	0.635793868153886\\
21	0.0778954719761771\\
22	0.0791250773291736\\
23	0.0365014080009551\\
24	0.113964497069553\\
};
\addlegendentry{bs 4};

\addplot [color=mycolor1,dotted,line width=1.5pt,forget plot]
  table[row sep=crcr]{%
1	0.96065025620466\\
2	0.96065025620466\\
3	0.96065025620466\\
4	0.96065025620466\\
5	0.96065025620466\\
6	0.96065025620466\\
7	0.96065025620466\\
8	0.96065025620466\\
9	0.96065025620466\\
10	0.96065025620466\\
11	0.96065025620466\\
12	0.96065025620466\\
13	0.96065025620466\\
14	0.96065025620466\\
15	0.96065025620466\\
16	0.96065025620466\\
17	0.96065025620466\\
18	0.96065025620466\\
19	0.96065025620466\\
20	0.96065025620466\\
21	0.96065025620466\\
22	0.96065025620466\\
23	0.96065025620466\\
24	0.96065025620466\\
};
\addplot [color=mycolor2,dotted,line width=1.5pt,forget plot]
  table[row sep=crcr]{%
1	0.449199811679103\\
2	0.449199811679103\\
3	0.449199811679103\\
4	0.449199811679103\\
5	0.449199811679103\\
6	0.449199811679103\\
7	0.449199811679103\\
8	0.449199811679103\\
9	0.449199811679103\\
10	0.449199811679103\\
11	0.449199811679103\\
12	0.449199811679103\\
13	0.449199811679103\\
14	0.449199811679103\\
15	0.449199811679103\\
16	0.449199811679103\\
17	0.449199811679103\\
18	0.449199811679103\\
19	0.449199811679103\\
20	0.449199811679103\\
21	0.449199811679103\\
22	0.449199811679103\\
23	0.449199811679103\\
24	0.449199811679103\\
};
\addplot [color=mycolor3,dotted,line width=1.5pt,forget plot]
  table[row sep=crcr]{%
1	0.34497791786781\\
2	0.34497791786781\\
3	0.34497791786781\\
4	0.34497791786781\\
5	0.34497791786781\\
6	0.34497791786781\\
7	0.34497791786781\\
8	0.34497791786781\\
9	0.34497791786781\\
10	0.34497791786781\\
11	0.34497791786781\\
12	0.34497791786781\\
13	0.34497791786781\\
14	0.34497791786781\\
15	0.34497791786781\\
16	0.34497791786781\\
17	0.34497791786781\\
18	0.34497791786781\\
19	0.34497791786781\\
20	0.34497791786781\\
21	0.34497791786781\\
22	0.34497791786781\\
23	0.34497791786781\\
24	0.34497791786781\\
};
\end{axis}
\end{tikzpicture}%
\hfill{}\input{img/PSNR_tecnick.tex}

\spaceBeforeLabel\vspace{2mm}\protect\caption{\label{fig:PSNR-per-image}Difference PSNR in dB of the reconstructed
pixels between the centroid adaption (\fsecentroid{}) and the unmodified
weighting function (\fse{}) for a dense loss pattern, Left: \textit{Kodak},
Right: \textit{Tecnick.}}

\spaceBelowFig\spaceBelowFig\spaceBelowFig\spaceBelowFig\spaceBelowFig
\end{figure*}
For the simulation, the loss patterns are generated using the following
commands in MATLAB
\begin{multline}
\hspace{-1mm}\mathtt{dense\_loss\text{\_}pattern}\hspace{-1mm}=\hspace{-1mm}\mathtt{imdilate(rand(img\text{\_}size)\hspace{-1mm}>\hspace{-1mm}0.98,}\\
\mathtt{strel(\text{'}square\text{'},8));}\label{eq:errmask1}
\end{multline}
\begin{multline}
\hspace{-1mm}\mathtt{sparse\_loss\text{\_}pattern}\hspace{-1mm}=\hspace{-1mm}\mathtt{imdilate(rand(img\text{\_}size)\hspace{-1mm}>\hspace{-1mm}0.95,}\\
\mathtt{strel(\text{'}square\text{'},8));}\label{eq:errmask2}
\end{multline}

Visual examples for the two loss patterns are shown in \Fig{\ref{fig:Eyecatcher}}.
Lost pixels are colored in black. The dense loss pattern in \eqref{eq:errmask1}
causes a loss of about 28\% of pixels and the sparse loss pattern
in \eqref{eq:errmask2} about 4\%, respectively. The challenging patterns
are very similar to the lost areas that occur in different applications.
We choose the dense pattern to evaluate the algorithms on larger lost
areas. 

For both loss pattern types, the simulations were repeated for 10
different loss patterns to obtain significant results and to avoid
special effects from a single specific pattern. For each loss pattern,
the PSNR is computed on the reconstructed pixels only. Then, for each
loss pattern, the average is computed for the 10 different loss patterns.
\Fig{\ref{fig:PSNR-per-image}} shows the results for the \textit{Kodak}
and the \textit{Tecnick} database for the dense loss pattern. Whenever
the difference is positive, \fsecentroid{} obtains a better reconstruction
result. The dotted lines show the average result for each test set.
The corresponding values to the dotted lines are listed as diff in
the first two lines of \Tab{\ref{tab:averaged-PSNR-results}} together
with the absolute PSNR values. In the lower part, \Tab{\ref{tab:averaged-PSNR-results}}
contains the corresponding averaged results of the simulation using
the sparse loss pattern. Further, both tables contain results from
four recent reference methods from the literature, namely based on
kernel regression (CKR) and (SKR) \cite{kernelregression2007}, total
variation (TV) \cite{totalvariation2010}, and constraint lagrangian
shrinkage (\CSALSA{}) \cite{salsa2011}.

\begin{table*}
\protect\caption{\label{tab:averaged-PSNR-results}Averaged PSNR values in dB using
the dense loss pattern \eqref{eq:errmask1} in the upper part and
the sparse loss pattern \eqref{eq:errmask2} in the lower part.}
\spaceBeforeLabel

 \setlength\tabcolsep{4pt} 

\begin{center}
\begin{tabular}{cc|rrrl|rrr|rrr|rrr}
\toprule
& & CKR\cite{kernelregression2007} & SKR\cite{kernelregression2007} & TV\cite{totalvariation2010} & \CSALSA & \multicolumn{3}{c|}{bs 4} & \multicolumn{3}{c|}{bs 8} & \multicolumn{3}{c}{bs 16}\tabularnewline
& & & & & \cite{salsa2011} & \fse{}\cite{seiler2010} & \fsecentroid{} & & \fse{}\cite{seiler2010} & \fsecentroid{} & & \fse{}\cite{seiler2010} & \fsecentroid{} &\tabularnewline
& & & & & & & proposed & diff~~ &  & proposed & diff~~ &  & proposed & diff~~ \tabularnewline
\midrule
\midrule

\multirow{4}{*}{\begin{turn}{90} ~~{\scriptsize Dense Loss Pattern}\end{turn}} \fakebuchstabe{} \vspace{0.55mm} 
 & \textit{Kodak}  & 21.952 & 21.476 & 23.264 & 23.856 & 25.419 & \textbf{25.764} & + 0.345 & 24.998 & \textbf{25.447} & + 0.449 & 23.895 & \textbf{24.856} & + 0.961\tabularnewline


\fakebuchstabe{} \vspace{0.55mm}   & \textit{Tecnick}  & 23.667 & 23.058 & 25.394 & 25.258 & 28.096 & \textbf{28.165} & + 0.069 & 27.529 & \textbf{27.733} & + 0.204 & 26.570 & \textbf{27.215} & + 0.645\tabularnewline


\fakebuchstabe{} \vspace{0.55mm}  & \textit{Middleburry}  & 28.564 & 27.524 & 31.354 & 31.227 & 34.912 & \textbf{35.025} & + 0.113 & 34.225 & \textbf{34.481} & + 0.256 & 33.086 & \textbf{33.874} & + 0.788\tabularnewline


\fakebuchstabe{} \vspace{0.55mm}  & \textit{Arri}  & 26.582 & 26.151 & 26.966 & 28.610 & 31.189 & \textbf{32.357} & + 1.168 & 30.497 & \textbf{31.528} & + 1.031 & 28.775 & \textbf{30.543} & + 1.768\tabularnewline

\\[2ex]

\midrule
%

\multirow{4}{*}{\begin{turn}{90} ~~{\scriptsize Sparse Loss Pattern}\end{turn}} \fakebuchstabe{} \vspace{0.55mm}  & \textit{Kodak}  & 24.602 & 25.931 & 25.209 & 26.219 & 28.105 & \textbf{28.782} & + 0.677 & 27.480 & \textbf{28.586} & + 1.106 & 25.055 & \textbf{27.866} & + 2.811\tabularnewline


\fakebuchstabe{} \vspace{0.55mm} 

 & \textit{Tecnick}  & 27.618 & 29.552 & 28.851 & 28.674 & 31.901 & \textbf{32.162} & + 0.261 & 31.265 & \textbf{31.932} & + 0.667 & 29.224 & \textbf{31.144} & + 1.920\tabularnewline


\fakebuchstabe{} \vspace{0.55mm}  & \textit{Middleburry}  & 33.979 & 36.285 & 35.359 & 35.278 & 39.545 & \textbf{39.913} & + 0.368 & 38.777 & \textbf{39.634} & + 0.857 & 36.305 & \textbf{38.667} & + 2.362\tabularnewline


\fakebuchstabe{} \vspace{0.55mm}  & \textit{Arri}  & 29.051 & 31.332 & 28.601 & 33.086 & 34.875 & \textbf{37.648} & + 2.773 & 33.919 & \textbf{37.206} & + 3.287 & 29.815 & \textbf{35.936} & + 6.121\tabularnewline

\end{tabular}
\end{center}

\spaceBelowTab\spaceBelowTab\spaceBelowTab\spaceBelowTab\spaceBelowTab\spaceBelowTab
\end{table*}

Generally, all PSNR values increase by about 3~dB for the sparse
loss pattern as can be seen by comparing the upper and the lower part
of \Tab{\ref{tab:averaged-PSNR-results}}. The sparse loss pattern
causes less and above all smaller lost areas. From the latter, all
methods can profit. Comparing the achieved gains of \fsecentroid{}
to the unmodified \fse{}, the gains also increase for less and smaller
lost areas. The weighting function $\fsew$ is centered on the centroid
of all lost pixels within the block~$\fsearearec$. When there are
less lost areas, the case of more than one distinct lost area within
$\fsearearec$ occurs less often. When there is only one lost area
within $\fsearearec$, the $\fsew$ can be centered exactly on the
centroid of this one lost area.

The behavior of the results is similar for all datasets with exception
of \textit{ARRI}, where \fsecentroid{} obtains an additional gain
of about 1~dB for the dense loss pattern compared to the other databases.
For the sparse loss pattern and bs~16, a remarkable gain of 6.1~dB
is obtained compared to \fse{}. We further investigated this case
and found that images from the \textit{ARRI} database have a small
black border. Slightly wrong reconstruction values on this border
can cause a very big loss in PSNR. Compared to the other databases,
the extreme large gains of \fsecentroid{} and \CSALSA{} for \textit{ARRI}
mostly come from border effects. To further analyze the performance
disregarding the border, we additionally evaluated the reconstruction
quality, omitting a border of 16~pixels. Excluding the image border,
the gain of 6.1~dB shrinks to 1.8~dB which is in the same range
compared to the other databases. The centroid adaption does not only
lead to a better reconstruction of lost areas at image borders but
also can improve the performance within the image.

The achieved reconstruction gain grows with increasing size of the
block~$\fsearearec$. The reason is that the centroid of lost pixels
is computed within the block $\fsearearec$. So, for a small block
size, the center of the weighting function can only be moved in a
small range. Nevertheless, the influence on the reconstruction result
is remarkable keeping in mind that for bs~4 the center can move by
a maximum $1.5\cdot\sqrt{2}\approx2.1\text{ pixels}$ for the extreme
case that exactly only one pixel located in one of the corners of~$\fsearearec$
is lost. With a larger block $\fsearearec$, the center can move to
a larger extend. This explains the increasing influence on the results
with increasing block size.

\begin{figure}
 \setlength\tabcolsep{2pt}

\begin{center}

\begin{tabular}{cccc}
{\footnotesize{}Original} & {\footnotesize{}CKR\cite{kernelregression2007}} & {\footnotesize{}TV\cite{totalvariation2010}} & {\footnotesize{}\fse{}\cite{seiler2010} bs8 }\tabularnewline
 &  &  & \tabularnewline
\includegraphics[height=4.2cm]{img_sim/Kodak_small_15_ErrId9_ErrP1_y1_x340_orig} & \includegraphics[height=4.2cm]{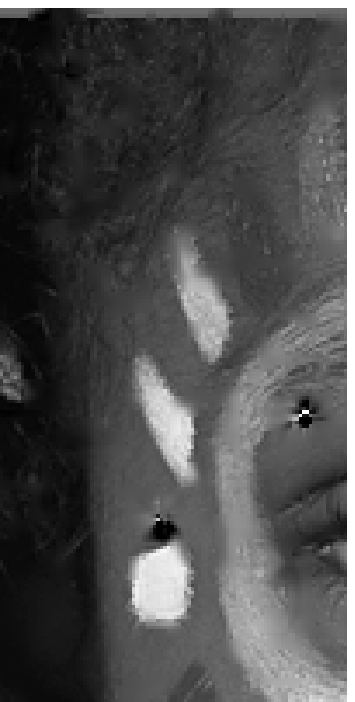} & \includegraphics[height=4.2cm]{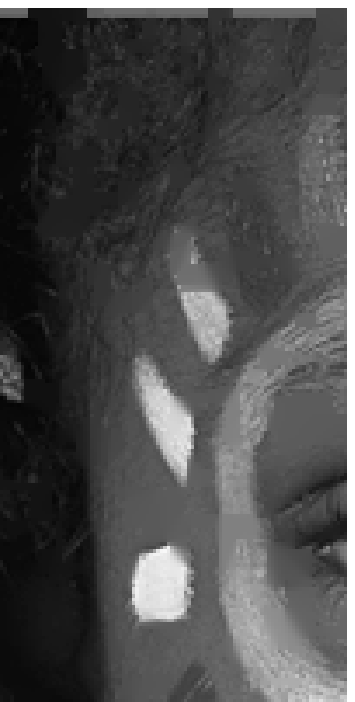} & \includegraphics[height=4.2cm]{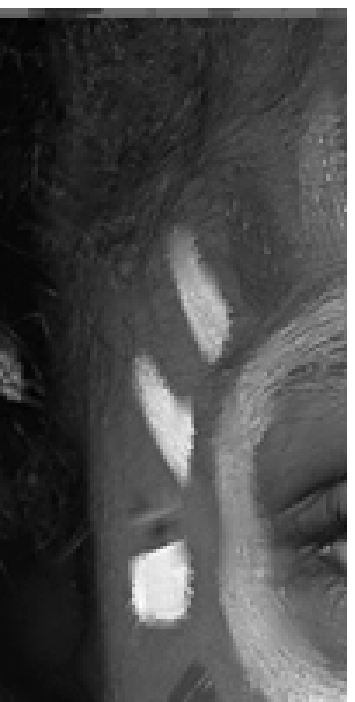}\tabularnewline
\includegraphics[height=4.2cm]{img_sim/Kodak_small_15_ErrId9_ErrP1_y1_x340_lost} & \includegraphics[height=4.2cm]{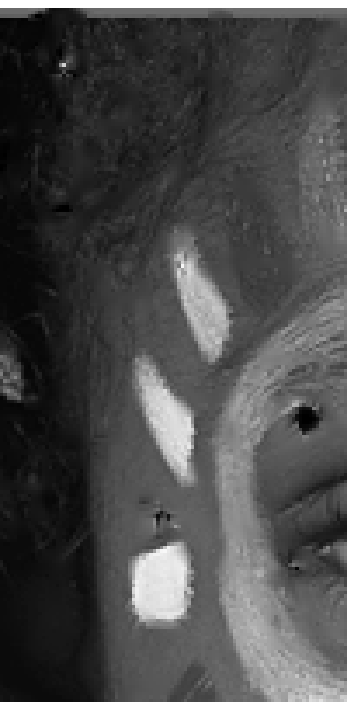} & \includegraphics[height=4.2cm]{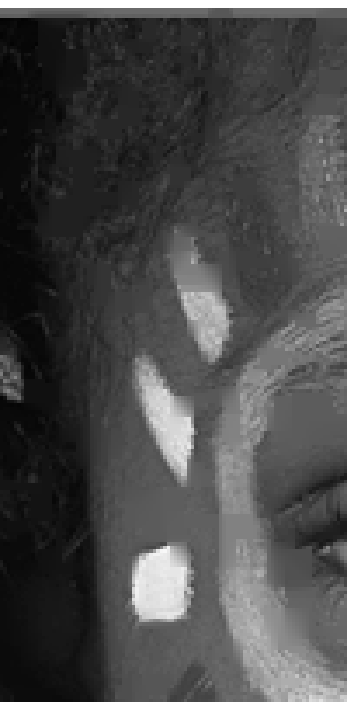} & \includegraphics[height=4.2cm]{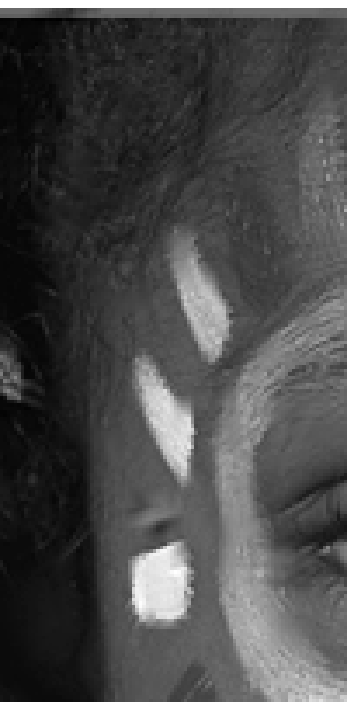}\tabularnewline
 &  &  & \tabularnewline
{\footnotesize{}Lost} & {\footnotesize{}SKR\cite{kernelregression2007}} & {\footnotesize{}\CSALSA{}\cite{salsa2011}} & {\footnotesize{}\fsecentroid{} bs8}\tabularnewline
 &  &  & {\footnotesize{}proposed}\tabularnewline
\end{tabular}

\end{center}

\spaceBeforeLabel

\protect\caption{\label{fig:Visual-inpainting-results}Visual reconstruction results
for a detail of image 15 of the \textit{Kodak} database using the
dense loss pattern. The lost areas are black in the lower left image.}

\spaceBelowFig\spaceBelowFig\spaceBelowFig\spaceBelowFig\spaceBelowFig\spaceBelowFig\spaceBelowFig\spaceBelowFig
\end{figure}

\Fig{\ref{fig:Visual-inpainting-results}} shows reconstruction results
for visual comparison. When pixels are lost in smooth image regions,
an adaption of the weighting function has no real influence on the
reconstruction result. In structured regions, and especially when
lost areas occur at edges, the proposed centroid adaption of the weighting
function is advantageous. The improved reconstruction performance
of \fsecentroid{} at the image border can be seen best for the gray
border at the top of the images.

\section{Conclusion}

\label{sec:Conclusion}In this paper, we consider the challenging
reconstruction of arbitrarily shaped lost areas in images. We propose
\fsecentroid{}, a centroid adaption of the weighting function of
\fse{}, to address the arbitrary shape of lost areas. Over our large
test set, \fsecentroid{} consistently improves the reconstruction
quality. The treatment of lost areas at image borders is improved.
With the proposed adaption, the reconstruction performance of \fse{}
can be improved by 1.29~dB of PSNR on average for arbitrarily shaped
lost image areas. .

Further work aims at optimizing the reconstruction, i.e., only one
distinct lost area within the currently considered block is reconstructed
at once. We also aim at an adaption of the shape of the weighting
function.

\section*{Acknowledgment}

The authors would like to thank Eduard Schön and Nils Genser for their
valuable assistance. Further, we gratefully acknowledge that this
work has been supported by the Deutsche Forschungsgemeinschaft (DFG)
under contract number \mbox{KA~926/4-2}.

\bibliographystyle{IEEEtran}
\bibliography{bib/bib_strings_short,bib/bibliography}


\end{document}